\documentstyle[aps,eqsecnum,multicol]{revtex}
\begin{document}

\draft
\title{Exact calculation of multifractal exponents of the critical
wave function of Dirac fermions in a random magnetic field}
\author{Horacio E.~ Castillo}
\address{Department of Physics, University of Illinois at
Urbana-Champaign, Urbana, IL 61801-3080}
\author{Claudio de C.~ Chamon and Eduardo Fradkin}
\address{Department of Physics, University of Illinois at
Urbana-Champaign, Urbana, IL 61801-3080 \\ and Institute for
Theoretical Physics, University of California at Santa Barbara, Santa
Barbara CA 93106-4030}
\author{Paul M.~ Goldbart}
\address{Department of Physics, University of Illinois at
Urbana-Champaign, Urbana, IL 61801-3080}
\author{Christopher Mudry}
\address{Department of Physics, Massachusetts Institute of Technology,
Cambridge, Massachusetts 02139 \\ and Institute for
Theoretical Physics, University of California at Santa Barbara, Santa
Barbara CA 93106-4030}

\date{September 16, 1997}

\maketitle

\begin{abstract}

The multifractal scaling exponents are calculated for the critical
wave function of a two-dimensional Dirac fermion in the presence of a
random magnetic field. It is shown that the problem of calculating the
multifractal spectrum maps into the thermodynamics of a static
particle in a random potential. The multifractal exponents are simply
given in terms of thermodynamic functions, such as free energy and
entropy, which 
are argued to be self-averaging in the thermodynamic limit.
These thermodynamic functions are shown to coincide exactly with those
of a Generalized Random Energy Model, in agreement with previous results
obtained using Gaussian field theories in an ultrametric space.

\end{abstract}



\begin{multicols}{2}

\section{Introduction}

In recent years it has become clear that the wave functions of noninteracting 
disordered systems 
at a continuous 
metal-insulator transition have multifractal scaling properties. As opposed to a 
simple fractal, 
these statistical self-similar wave functions cannot be described by a single 
fractal
dimension, but instead an infinite set of scaling exponents is needed.
Indeed, such families of scaling exponents have been obtained for the critical 
wave function at
a localization-delocalization transition 
within several different frameworks, including 
perturbative 
renormalization-group treatments on replicated\cite{Wegner 1980} and 
supersymmetric\cite{Falko 1995} nonlinear sigma models, as well as numerical 
simulations\cite{Janssen 1994}. Although these results seem to provide 
sufficient evidence for 
multifractality at the metal-insulator transition, none of them allows one to 
understand the full 
spectrum of multifractal exponents. This is so because nonperturbative 
techniques are needed to probe the full spectrum.

Given the present state of affairs of this problem it would 
be highly desirable 
to have an exactly 
solvable system that exhibits a multifractal wave function. 
In recent years it 
has been shown 
that the Dirac equation in random fields in two-spatial 
dimensions (2D) is 
actually an example of such a system. 
Moreover, it has also become clear that 
the random 
Dirac Hamiltonian in 2D describes the universality class of the metal-insulator 
transition 
at a half-filled Landau level of disordered {\it nonrelativistic} 
noninteracting electrons 
in a very high magnetic field\cite{Fisher-Fradkin 1986,Ludwig 1994}. This 
surprising result 
has been recently established by direct derivation \cite{Ho 1996} of 
the random Dirac equation 
from the Chalker-Coddington model of percolating edge states
\cite{Chalker 1988}.

Recently, {\em exact} results for the {\em full} multifractal spectrum 
have been obtained for 
Dirac fermions interacting with a random magnetic field \cite{CMW}. 
These results were 
obtained by mapping the problem to a Gaussian field theory in an ultrametric 
space (a Cayley tree).
The calculation of multifractal scaling properties was then reduced to the 
computation of 
thermodynamic functions of a special generalized random energy model (GREM), 
which are known 
exactly\cite{Derrida 1980,Derrida 1985,Derrida 1988,Buffet 1993}.  
The field theory in the 
localization problem is defined in Euclidean space. Nevertheless, 
it was argued that 
as the exponents are a measure of a global property, 
the difference between the ultrametric and the 
Euclidean metric would not alter the 
results. Moreover, a phase transition that occurred in the ultrametric 
model\cite{Derrida 1988} 
also seems to be present in the original Euclidean problem, as evidenced  by 
analytical arguments 
and by Monte Carlo simulations\cite{CMW}.

It has also been pointed out that there exists a deep connection
between the multifractal spectrum for the critical wave function of a Dirac 
fermion in a random 
magnetic field and the spectrum of primary fields in non-unitary
conformal field theories with vanishing Virasoro conformal charge\cite{Mudry 
1996,Caux 1996}
on the one hand, and Liouville field theory\cite{Kogan 1996}
on the other. 
It is then tempting to speculate that there must be a 
counterpart to the 
freezing transition characterizing GREM in these quantum field theories. 

In the present paper we show that the exact results of
Ref.~\cite{CMW} on the multifractal spectrum and, in particular, the
existence of a phase transition, can be obtained completely
within the framework of the Gaussian field theory in two-dimensional Euclidean
space. 
The key idea of our approach is the introduction of a quantity 
$\Omega(E)$ that counts the number of points 
where the critical wave function has amplitude (or height) 
$\Psi \propto e^{-E/2}$  
in a disk of radius $L$ 
(suitably discretized as a lattice of spacing $a$).
We show that this quantity
is directly related to the microcanonical density of states of the GREM 
and which we will also call the density of states. The advantage of the 
approach that we present in this work is that it is direct and it does 
not rely on the use of either replicas or supersymmetry. 
Our results are essentially rigorous, except for a 
conjecture, which we believe to be true, 
about the self-averaging property of the
probability distribution of the entropy 
$\ln \Omega(E)$ in the thermodynamic limit 
$N \to \infty$. We also explain in detail why the Gaussian field theories
in both Euclidean and ultrametric spaces give exactly the same results.

This paper is organized as follows. In Sec.~\ref{sec:model} we introduce the 
model and the quantities of main interest: 
the multifractal exponents $\tau(q)$ of the critical wave function,
its Legendre transform $f(\alpha)$ and the density of states $\Omega(E)$. 
Here we draw an analogy between the computation of 
the inverse participation ratio of order $q$ of the wave function and the 
partition function for all spatial configurations of a 
static particle 
in a random potential 
$\phi({\bf x})=-\ln |\Psi({\bf x})|$. 
$\Omega(E)$ is the microcanonical density of states of this equivalent problem.
In Sec.~\ref{sec:thermo} we derive the thermodynamic properties of the 
equivalent system. Here we calculate the average density of states 
and use it to show that the equivalent 
problem has a phase transition at a critical 
``energy" determined by the width $g$ of the probability 
distribution of the random vector potentials.
We show that the thermodynamic functions of the 
equivalent problem are 
exactly those of the GREM. 
In particular, we show that, in terms of the critical wave function, the phase 
transition of the GREM  represents 
the onset of the regime where the probability distribution of 
the wave function is {\it undersampled} in a given 
discretization of the plane.
In Sec.~\ref{sec:spectrum} we use these results to derive the exact 
form of the functions $f(\alpha)$ and $\tau(q)$. Our results agree completely 
with the analysis 
given in Ref.~\cite{CMW}. 
Section \ref{sec:conclusions} is devoted to the 
conclusions. Technical 
details of our calculations are given in the appendices.

\section{Model}
\label{sec:model}

We consider the problem of a massless Dirac fermion moving on a plane
and interacting with a static random magnetic field normal to the plane
\cite{Fisher-Fradkin 1986,Ludwig 1994,Nersesyan 1995,Mudry 1996,Ho 1996,Wen-Ho}.
In this model, the wave functions are localized for all energies other
than the critical energy $E=0$, 
at which the wave function is multifractal\cite{Ludwig 1994,Mudry 1996}. 
This model thus describes a metal-insulator transition in two dimensions.

The Dirac Hamiltonian in random and static 
{\it vector potentials} in two space dimensions 
is
\begin{equation}
H=\sigma_\mu \; \left[i v_F  \partial_\mu- A_\mu({\bf x}) \right].
\label{eq:dirac}
\end{equation}
For convenience we will set the Fermi velocity $v_F$ to 
unity from now on.
This operator acts on the space of normalizable (in a finite area)
two-component spinor states 
$\Psi_\alpha({\bf x})$, with $\alpha=1,2$. 
In Eq.~(\ref{eq:dirac}) 
{\boldmath ${\sigma}$} denotes a two-component 
vector of two $2 \times 2$ Pauli matrices, 
which we take to be $\sigma_1$ and $\sigma_2$ respectively, with 
$\mu=1,2$ being the two orthogonal directions on the plane. 
The probability distribution of the random vector potential 
${\bf A}({\bf x})$ will be specified below. 
In principle, 
other sorts of random 
fields, such as random mass and random chemical potential, 
are also allowed in a general situation\cite{Fisher-Fradkin 1986}.

If only random vector potentials are allowed, 
$E=0$ is an exact eigenenergy for all realizations of the disorder and
the corresponding wave functions can be determined  exactly 
\cite{Aharonov 1979}. 
Indeed, let $\Psi_0$ be a two-component spinor with energy $E=0$, 
 {i.e.}, $H \Psi_0=0$. Then by means of a combination of 
chiral and gauge transformations, parametrized by the fields 
$\phi({\bf x})$ and $\chi({\bf x})$, respectively,
\begin{equation}
\Psi_0({\bf x})=e^{-\phi({\bf x}) \sigma_3-i \chi({\bf x})} \; \eta({\bf x}),
\label{eq:chiral}
\end{equation}
the eigenvalue equation reduces to the requirement that the two component 
spinor $\eta({\bf x})$ 
satisfies the nonrandom Dirac equation
\begin{equation}
i\sigma_\mu \partial_\mu \eta({\bf x})=0.
\label{eq:nonrandom}
\end{equation}
Here, the chiral ``angle" $\phi ({\bf x})$ and the gauge transformation 
$\chi({\bf x})$ must be chosen 
to solve Eq.~(\ref{eq:A}) below.

In this work we will only be interested in the multifractal properties of the 
amplitudes of the wave functions. These properties involve only the magnitude 
of the wave functions and are independent of their 
phases, and hence are gauge 
invariant properties.
However, it should be stressed that the decomposition of Eq.~(\ref{eq:A})
is only valid if the total magnetic flux threading the disk always vanishes, 
an assumption 
that we will implement by an appropriate choice for the probability 
distribution of the vector potential ${\bf A(x)}$
\cite{footnote on topo trivial}.
The remaining degrees of freedom carried by the spinor $\eta({\bf x})$
then span a two-dimensional Hilbert space. 
Here we will {\it choose} the spinor $\eta({\bf x})=(1,0)$ for convenience. 
It is worth noting that any choice of spinor breaks 
the chiral symmetry generated by $\sigma_3$. 
This procedure makes sense if we think of switching on an average uniform 
magnetic field and then taking it to zero. 
Indeed, a magnetic field selects a state with 
unique chirality determined by the sign of the magnetic field. 
In the context of the Chalker-Coddington model 
the choice of spinor is thus the equivalent of the choice of the chirality 
of the edge current for a system on an open geometry such as a disk. 

With the above considerations we write the $E=0$ wave function as
\begin{equation}
\psi({\bf x}) = e^{-\phi({\bf x})} ,
\label{eq:psi_phi} 
\end{equation}
and drop the constant spinor $\eta({\bf x})=(1,0)$ altogether. 
The random vector potential and magnetic field are given by:
\begin{eqnarray}
A_{\nu}({\bf x}) & = & \epsilon_{\nu \rho} \partial_{\rho} \phi({\bf x}) 
+ \partial_{\nu} \chi({\bf x}), 
\label{eq:A}\\
B({\bf x}) & = & - \nabla^{2} \phi({\bf x}).
\label{eq:AB}
\end{eqnarray}
Notice that the gauge degrees of freedom enter through $\chi$ and not
$\phi$, and any phase in the wave function can be eliminated by a gauge
transformation.
Finally, we assume a Gaussian distribution of magnetic fields \cite{appA}
as follows:
\begin{equation}
P[\phi({\bf x})] \propto e^{-\frac{1}{2 g} \int d^{2} {\bf x}
( \nabla \phi({\bf x}) )^2 }.
\label{eq:distrib}
\end{equation}
where $g$ is the width of the probability distribution and plays the role of a 
coupling constant in this problem. 
One verifies that, 
in the thermodynamic limit, the uniform magnetic field does indeed vanish 
for all realizations of the disorder. 
As a technical point, it is understood that, among all fields
$\phi({\bf x})$ that differ from each other by a uniform value, only
one representative is counted in the disorder average over $\phi$ [for
example, of all possible uniform fields, only the field $\phi({\bf
x}) = 0$ is counted].  In this way, the ambiguity in the many to one
relation between $\phi({\bf x})$ and $\bf A(x)$ in Eq.~(\ref{eq:A}) is
removed (see Appendix \ref{sec:GF}).

The multifractal nature of 
the wave function can be probed through the
moments of the probabilities $p_{\bf x}$ obtained from the normalized
wave function $\Psi({\bf x})$. 
As we anticipated above, we shall use a lattice regularization
with a microscopic cutoff distance $a$ (box size), and macroscopic
system size $L$. 
There are thus $L^2/a^2$ sites. The moments of
$p_{\bf x}$ are the inverse participation ratios ${\cal P}(q,a/L)$:
\begin{eqnarray}
{\cal P}(q,{a\over L}) & \equiv & \sum_{{\bf x}} p^{q}_{\bf x}
\nonumber\\
& \equiv & \sum_{\bf x} | \Psi({\bf x}) |^{2q}
\nonumber\\
& \equiv & \frac{ \sum_{{\bf x}} | \psi({\bf x}) |^{2q} }
{ (\sum_{{\bf x}} | \psi({\bf x}) |^{2} )^q } 
\nonumber\\
& = & \frac{\sum_{\bf x} e^{-2 q \phi({\bf x})}}
{\left\{\sum_{\bf x} e^{-2 \phi({\bf x})}\right\}^{q}}.
\end{eqnarray}
The multifractal exponents $\tau(q)$ for a given wave function
are then defined by\cite{Hentschel 1983,Grassberger 1983}
\begin{eqnarray}
\tau(q)& \equiv &D(q)(q-1) 
\nonumber\\
& \equiv &
\lim_{\frac{a}{L} \rightarrow 0} \frac{1}{\ln(\frac{a}{L})}
\ln\left(\sum_{{\bf x}} p^q_{\bf x}\right) 
\nonumber\\
& = & \lim_{\frac{a}{L} \rightarrow 0} 
\frac{\ln{{\cal P}(q,a/L)}}{\ln(\frac{a}{L})}.
\label{eq:def_tau}
\end{eqnarray}
We will show later that $\tau(q)$ is self-averaging, {i.e.}, 
that for any
realization of the disorder $\tau(q) =  \langle \tau(q) \rangle $, where $ 
\langle \cdots \rangle $
denotes the average over the distribution of random magnetic fields of
Eq.(\ref{eq:distrib}).

An equivalent way of describing the multifractal properties of 
a {\em given} wave function is through the scaling exponents $\alpha_{\bf x}$, 
defined by\cite{Mandelbrot 1974,Frisch 1985,Halsey 1986}
\begin{equation}
p_{\bf x} \sim \left(\frac{a}{L}\right)^{\alpha_{\bf x}}.
\label{eq:p_alpha}
\end{equation}
The number of lattice points at which the exponent $\alpha_{\bf x}$ takes values 
between $\alpha'$ and $\alpha' + d\alpha'$ defines the functions
$\rho$ and $f$:
\begin{equation}
d\alpha' \rho(\alpha') \left(\frac{a}{L}\right)^{-f(\alpha')}.
\label{eq:f_def}
\end{equation}
As is well known\cite{Frisch 1985,Halsey 1986}, 
the two sets of exponents $\tau(q)$ and $f(\alpha)$
are related by a Legendre
transformation:
\begin{eqnarray}
&&\alpha=\frac{d\tau(q)}{dq},
\label{eq:alpha}\\
&&f(\alpha)=\alpha q -\tau(q).
\label{eq:f_alpha_legendre}
\end{eqnarray}

It is possible in general to map the quantities that describe a given
multifractal wave function into thermodynamic quantities 
\cite{Jensen 1978}:
$q$ maps into an inverse
temperature $\beta$, $\tau(q)$ into a free energy $f_{0}(\beta)$,
$\alpha$ into an internal energy $e$, and $f(\alpha)$ into an
entropy $s(e)$. 
In our case, however, the equivalence is more evident due
to the particular form of the wave function of Eq.~(\ref{eq:psi_phi}).
We map our problem into the statistical mechanics of
a single particle in a lattice of spacing $a$ and size $L$ with a
random site potential $V({\bf x}) = 2\phi({\bf x})$, in the static
limit ({i.e.}, for the hopping matrix element equal to zero). For this model, the 
random
canonical partition function for a particular realization of the
disorder reads:
\begin{equation}
Z(\beta) \equiv \sum_{\bf x} e^{-\beta V({\bf x})},
\label{eq:def_can}
\end{equation}
where the role of the random energies is played by 
the values that the disorder potential $V({\bf x}) = 2\phi({\bf x})$ takes. 
For each disorder realization, the free energy of the system is
\begin{equation}
F(\beta) \equiv -\frac{1}{\beta}\ln{Z(\beta)}.
\label{eq:free_edef}
\end{equation}
In this system, the number of thermodynamic degrees of freedom is
\begin{equation}N =
\ln{\left(\frac{L}{a}\right)^{2}},
\end{equation}
and the number of energy levels is 
\begin{equation}
e^{N} =
\left(\frac{L}{a}\right)^{2}. 
\end{equation}
This is similar to a system of $N$ spins with 
$S = \frac{1}{2}$, where one has $2^{N}$ states. 
Thus, the intensive free energy is defined as 
\begin{equation}
f_{0}(\beta) \equiv \frac{F(\beta)}{N}.
\end{equation}
[The subscript $0$ is introduced to avoid confusion with the spectrum 
$f(\alpha)$.]

In this problem it turns out that it is also useful to count states
directly, which naturally leads us to define a microcanonical partition
function (or density of states) for each random field configuration:
\begin{equation}
\Omega(E) \equiv \sum_{\bf x} \delta_{W}\left[E - V({\bf x})\right].
\label{eq:def_micro}
\end{equation}
Here $\delta_{W}(E)$ counts the number of states in a region of width
$W$ around $E$.
One can choose for it, {e.g.}, either a top hat ({i.e.}, a
product of step functions) $\delta_{W}(E) = \theta(E-W/2) \:
\theta(W/2-E),$ or a smooth function such as 
\begin{equation}
\delta_{W}(E) \equiv
e^{-\frac{E^{2}}{2 W^2}}.
\end{equation} 
The microcanonical and canonical partition
functions are then related by Laplace transformation:
\begin{equation}
Z(\beta) =  \int {d{E}\over W}\; \Omega(E)\ e^{-\beta E}.
\label{eq:can_micro}
\end{equation}
Here, as is usual in statistical mechanics, $W$ has to be taken as
small as possible but still larger than the average level spacing
$\Delta$. As we show later, in the present case this translates into the
condition:
\begin{equation}
1 \gg W \gg \Delta 
 = 4 \sqrt{\frac{g}{2 \pi}}\ 
\frac{ \ln \left(\frac{L}{a}\right)^2 }{\left(\frac{L}{a}\right)^2}.
\label{eq:spacing}
\end{equation}
{}From the microcanonical partition function one can obtain the 
total entropy:
\begin{equation}
S(E) \equiv \ln{\Omega(E)},
\label{eq:S(E)}
\end{equation}
and again it will be convenient to define the entropy and energy per 
thermodynamic degree of freedom: 
\begin{equation}
s \equiv S/N,\quad e \equiv E/N.
\end{equation}

Having established the thermodynamic dictionary, the normalization 
factor for the wave function and  
the inverse participation ratios ${\cal P}(q,a/L)$ can be written in 
terms of the partition function:
\begin{eqnarray}
\Psi({\bf x}) & = & \frac{e^{-\phi({\bf x})}}{\sqrt{Z(1)}},
\label{eq:Psi_normalization} \\
{\cal P}(q,{a\over L}) & = & \frac{Z(q)}{{Z(1)}^{q}}.
\end{eqnarray}
In turn, the function $\tau(q)$ can be simply expressed in terms of the
free energy at the inverse temperatures $\beta = q$ and $\beta = 1$:
\begin{eqnarray}
\tau(q) &=& -q \lim_{\frac{a}{L} \rightarrow 0} 
\frac{F(q) - F(1)}{\ln(\frac{a}{L})}\nonumber\\
&=& 2 q \lim_{\frac{a}{L} \rightarrow 0} 
\left[f_0(q)-f_0(1)\right].
\label{eq:tau_can}
\end{eqnarray}
Similarly, from the definition of $\alpha_{\bf x}$
[see Eq.~(\ref{eq:p_alpha})] and the value of the wave function
[Eq.~(\ref{eq:Psi_normalization})], we see that
\begin{equation}
\alpha_{\bf x} =
2\left[ \frac{V({\bf x})}{\ln\left(\frac{L}{a}\right)^2} -  f_0(1) \right].
\end{equation}
Thus, comparing the definitions of $f(\alpha)$ [Eq.~(\ref{eq:f_def})] and
the entropy, one verifies that
\begin{eqnarray}
&&\alpha = 2[e-f_0(1)], 
\nonumber\\
&&f(\alpha) = 2 s(e).
\label{eq:f_s}
\end{eqnarray}
By using the relation between $\tau(q)$ and $f_{0}(\beta)$, and 
between $f(\alpha)$ and $s(e)$ one recovers the well known result 
that the Legendre transformation in the language of multifractal 
exponents [Eq.~(\ref{eq:f_alpha_legendre})] is equivalent to the thermodynamic
relation:
\begin{equation}
f_{0}(\beta) = e - \frac{s(e)}{\beta}.
\label{eq:thermo_legendre}
\end{equation}

\section{Thermodynamics of the model}
\label{sec:thermo}

In this section, we are going to show that the probability to find energy 
levels outside a particular window of energy $[-e_c,e_c]$ 
vanishes in the thermodynamic limit. This is the most important 
result of our work.

The key step in our argument
is our estimate for the disorder average of 
the microcanonical partition function, which relies crucially on the choice
of a distribution for the disorder with a variance depending
logarithmically on spatial separation.

The simplest way to study the thermodynamic properties of the system
is to obtain the disorder average of the microcanonical partition function:
\begin{eqnarray}
 \langle \Omega(E) \rangle  & = & 
\int {\cal D}{V({\bf x})} P[V({\bf x})] 
\sum_{\bf x} \delta_{W}\!\left[E - V({\bf x})\right] 
\nonumber\\
& = & \int {\cal D}{\phi({\bf x})} P[\phi({\bf x})] 
\sum_{\bf x} \delta_{W}\!\left[E - 2\phi({\bf x})\right] 
\nonumber\\
& = & W \sum_{\bf x} 
\frac{ \exp \left\{ - E^{2}/2\left[W^2+4G({\bf x},{\bf x})\right] \right\}  }
{ \sqrt{W^2+4G({\bf x},{\bf x})} } 
\label{eq:micro_sum} 
\\
& \approx & W \left(\frac{L}{a}\right)^{2} 
\frac{ e^{-E^{2}/\left[\frac{4g}{ \pi} \ln\left(\frac{L}{a}\right)\right]} }
{ \sqrt{ \frac{2g}{ \pi} \ln\left(\frac{L}{a}\right) } }.
\label{eq:micro_E}
\end{eqnarray}
Here,  $G({\bf x},{\bf y})$
is the Green function \cite{appA} for the field
theory defined by the action of Eq.~(\ref{eq:distrib}), 
with a short distance cutoff length given by $a$. Its form in 
the short distance limit is 
\begin{equation}
G({\bf x},{\bf y}) = -{g\over4\pi}
\ln\left({|{\bf x}-{\bf y}|^2 + a^2 \over L^2}\right) . 
\end{equation}
On the last line of
Eq.~(\ref{eq:micro_E}), 
we have neglected the width $W$ as compared
to terms of order $\ln\left(\frac{L}{a}\right)$.

Alternatively,
in terms of intensive quantities, the disorder averaged number of states in 
an energy interval of width $w = W/N$ around $e=E/N$ is:
\begin{eqnarray}
 \langle \Omega(e) \rangle  & \approx & 
w \sqrt{\frac{ \ln\left(\frac{L}{a}\right)}{g} }\
e^{-2 \ln\left(\frac{L}{a}\right)
\left(\frac{e^{2}}{4(g/2\pi)} - 1\right)} 
\nonumber\\
& = & w \sqrt{\frac{N}{2g} }
\ e^{N\left(1-{\frac{e^{2}}{4(g/2\pi)}}\right)}.
\label{eq:micro_e}
\end{eqnarray}
This means that in the thermodynamic limit ($N \rightarrow \infty$),
$ \langle \Omega(e) \rangle $ goes to zero for $|e| > 2\sqrt{{g}/{2 \pi}}$ 
and
diverges exponentially with the number of degrees of freedom for 
$|e|< 2\sqrt{{g}/{2 \pi}}$. This behavior for the average number of
states $ \langle \Omega(e) \rangle $ indicates that some sort of transition 
should
occur at the critical energy $|e|=e_{c}$ where
\begin{equation}
e_{c} \equiv 2\sqrt{\frac{g}{2\pi}}.
\end{equation}
To show that, indeed, this is a phase transition, we must
consider not the average $ \langle \Omega(e) \rangle $, 
but the number of states
$\Omega(e)$ for {\em a given realization}. 
Indeed, as shown in Appendix \ref{sec:some moments},
$\Omega(E)$ is a random variable with very strong fluctuations. 
Hence, we cannot rely, {\it a priori}, on any single
moment of $\Omega(E)$ to identify a phase transition.
Rather, we must use stronger probabilistic arguments to show that
$e_c$ is indeed {\it the critical energy of interest}.
We do so below by considering the
two regimes $|e| > e_{c}$ and $|e| < e_{c}$ separately.

\subsection{$|e| \geq e_{c}$}

In this region the thermodynamic limit of the microcanonical partition
function is $ \langle \Omega(e) \rangle \rightarrow 0$.
Alternatively, we may say that for a large enough system size $N$, the
average number of states $ \langle \Omega(e) \rangle \ll 1$ in this
region. If we naively try to obtain the entropy as $\ln \langle
\Omega(e) \rangle $, we would find that it becomes negative for $|e| >
e_{c}$. This is {\em not} correct; the entropy must be defined for
{\em each and every} realization of the disorder. That means we must
focus on $S(e)=\ln \Omega(e)$ for separate realizations.

We are going to show now that the probability of finding {\em any} 
state at all with $|e| > e_{c}$ vanishes in the thermodynamic limit.
To do that, we define the random variable 
$\Omega_{>}(e)$ that counts the number of states with energies $e'$ 
such that $|e'| > e$:
\begin{equation}
\Omega_{>}(e) \equiv  \int_{-\infty}^{-e} d e' \frac{\Omega(e')}{w}
+ \int_{e}^{\infty} d e' \frac{\Omega(e')}{w}.
\label{eq:Omega_plus_def} 
\end{equation}
Due to the fact that $\Omega_{>}(e)$ is either positive or zero, 
we can bound the probability $P\{ \Omega_{>}(e) \ge 1 \}$ of finding at 
least one state with $|e'| > e$ by the average of $\Omega_{>}(e)$ \cite{bound}:
\begin{equation}
P\{ \Omega_{>}(e) \ge 1 \}  \leq   \langle  \Omega_{>}(e)  \rangle .  
\label{eq:Omega_plus_bound}
\end{equation}
Equation (\ref{eq:Omega_plus_bound}) is very general since it applies to 
any random microcanonical partition function. However, this 
inequality becomes very powerful when combined with our estimate 
Eq.~(\ref{eq:micro_e}) for $ \langle \Omega(e) \rangle $.
Indeed, with the help of
\begin{eqnarray}
{ \langle \Omega_{>}(e) \rangle }\! 
& = & \frac{2}{w} \int_{e}^{\infty} d e'  \langle \Omega(e') \rangle  
\label{eq:Omega_plus_e}\\
& = & \!
{\frac{1}{\sqrt{\pi N} (e/e_c)}}
e^{-N\!\left[\left(e/e_c\right)^2 - 1 \right]} 
\left[1 - {\cal O}\left(\frac{e_c}{N e}\right)\right],
\nonumber
\end{eqnarray}
and our upper bound Eq.~(\ref{eq:Omega_plus_bound}), we see that 
$P\{\Omega_>(e_c)\geq1\}$ vanishes in the thermodynamic limit $N\rightarrow\infty$.
We conclude that, {\em for a given realization}, the energy
levels will fall within the interval $|e| \le e_c$ with probability 1 in the
thermodynamic limit.

It is now clear how to estimate the average level spacing
$\Delta$ in Eq.~(\ref{eq:spacing}).
We simply multiply $2e_c$ by the ratio of the number 
$N$ of thermodynamic degrees of freedom to the number $e^N$ of energy levels:
\begin{equation} 
\Delta = \frac{2N e_{c}}{e^N} = 4 \sqrt{\frac{g}{2 \pi}} 
\frac{\ln\left(\frac{L}{a}\right)^2}{\left(\frac{L}{a}\right)^2}.
\label{eq:delta_N}
\end{equation}

Finally, we can bound from above the probability for the density of states
$\Omega(e)$ to be nonzero.
For a given realization, $\Omega(e)$ counts states in an energy
interval $w = W/N$ around $e$. Therefore $\Omega(e)$ must be a positive integer 
\cite{Heaviside},
in contrast with the average $ \langle \Omega(e) \rangle $. In the same way as 
before, 
the probability of having a nonzero microcanonical 
partition function is bounded from above
by a number that goes to zero in the thermodynamic limit \cite{bound}:
\begin{equation}
P\{ \Omega(e) > 0 \} = P\{ \Omega(e) \ge 1  \} \leq   \langle \Omega(e) \rangle 
.
\label{eq:P_Omega_bound}
\end{equation}
The entropy is therefore {\em not} defined for $|e| > e_c$.

One can get some intuition for these results with one word: {\em
undersampling}. Consider $\Omega(e)$ for an individual realization
as the histogram of the number of states per energy interval.
 One can think of $\Omega(e)$ as the product of two 
factors: one is a constant
that counts the total number of states in the system, and the other is
the probability (normalized to $1$) for {\em one} state to fall within
a given energy interval. In the present case the first factor has the value
$e^N$ and the second is Gaussian with a width that grows linearly with
$N$. In other words, the number of data points we have is $e^N$, which
is not enough to sample the tails of the Gaussian distribution. If the
number of data points were $e^{N^\gamma}$, with $\gamma>1$, by taking
$N$ large enough we could sample the whole distribution, and all bins
in the histogram would contain a positive number of points.

\subsection{$|e| < e_{c}$}

In this case we can define the entropy for a given realization
$S(e)=\ln \Omega(e)$ [recall Eq.~(\ref{eq:S(E)})]. We will now show that
the entropy, for all realizations of the disorder, has a common upper 
bound that scales linearly with system size. 
Furthermore, we will also show that 
the probability to find a realization with entropy 
{\em less or equal} 
than $\ln \langle \Omega(e)\rangle$ is 
{\it equal to one} in the $N \to \infty$ limit. 
The arguments go as follows.

According to our definition Eq.~(\ref{eq:def_micro}), 
$\Omega(e)$ counts the number
of energy levels in some energy window. That number cannot be larger than 
the total number of energy levels $e^N=L^2/a^2$. Consequently, $\ln\Omega(e)$
is bounded from above by $N$. Incidentally, no such bound holds
for $\ln Z(\beta)$ for individual realizations of the disorder.

Next, we introduce the probability to find the intensive entropy 
$s=(1/N)\ln\Omega(e)$ in a given interval $(s_1,s_2)$ by
\begin{equation}
P_N(s_1 \leq s \leq s_2) \equiv 
\int_{s_1}^{s_2} d \mu_N(s).
\label{eq:probab}
\end{equation}
Here, $d\mu_N(s)$ is the measure of the entropy for $N$ degrees of freedom.
We will make the following assumptions:
\begin{enumerate}
\item
The probability measure $d\mu_N(s)$ has a well defined thermodynamic limit:
\begin{equation}
\lim_{N\to \infty} \int_{s_1}^{s_2} d \mu_N(s)=
\int_{s_1}^{s_2} d \mu_\infty(s)
\label{eq:convergence}
\end{equation}
for all $s_1<s_2$.
\item
There 
exist two positive constants $A$ and $B$ such that
\begin{equation}
\langle \Omega(e)\rangle\equiv\int_{-\infty}^{+\infty} d\mu_N(s) e^{Ns} =
{\sqrt {N}} e^{A N+B}
\label{eq:large}
\end{equation}
for $N$ sufficiently large.
\end{enumerate}
Notice that we have proved the validity of the second hypothesis
and that $A$ and $B$ can both be 
read off from Eq.~(\ref{eq:micro_e}).
In particular,
\begin{equation}
A=1-{e^2\over e^2_c}=
\lim_{N\to \infty}{1\over N}\ln \langle \Omega(e) \rangle.
\end{equation}
We now choose $s_+>A$. By assumption, $s_+>0$ and 
\begin{eqnarray}
\langle\Omega(e)\rangle
&&\geq\int_{s_+}^{+\infty}d\mu_N(s) e^{Ns}
\nonumber\\
&&\geq\int_{s_+}^{+\infty}d\mu_N(s) e^{Ns_+}
\nonumber\\
&&\equiv e^{Ns_+} P_N(s_+\leq s).
\end{eqnarray}
We have thus established the upper bound,
\begin{equation}
P_N(s_+\leq s)\leq {\sqrt {N}} e^{(A-s_+) N+B},
\end{equation}
for all $s_+>A$ and for all sufficiently large $N$.
Since the thermodynamic limit is assumed to be well defined,
we conclude that
\begin{equation}
P_\infty(A< s)=0.
\label{eq:bound}
\end{equation}
In other words,
the probability to find the entropy density 
$(1/N) \ln \Omega(e) >
(1/N) \ln \langle \Omega(e) \rangle$ 
vanishes in the thermodynamic limit $N \to \infty$. 

If we could prove that the probability to find an intensive entropy
lower than $(1/N)\ln\langle\Omega(e)\rangle$ 
also vanishes in the thermodynamic limit,
we would have shown that the entropy density is indeed self-averaging 
({i.e.}, almost surely takes one and only one value),
and given by $\lim_{N\rightarrow\infty}(1/N)\ln\langle\Omega(e)\rangle$.
However, without further knowledge of the probability 
distribution for the entropy, it is not possible to prove the existence 
of a similar lower bound. In fact it is possible to construct a sequence of
measures which obey Eqs. (\ref{eq:convergence}) and (\ref{eq:large}) 
but is not self-averaging in the thermodynamic limit.
We nevertheless expect the entropy density to be self-averaging
for two reasons. First and in addition to our upper bound,
it is possible to estimate the ratio of the 
$n$-th moment of the
partition function $Z(\beta)$ to its mean raised to the power $n$.
This is done in Appendix \ref{sec:some moments}. 
This estimate suggests that the field theory
computations of the multifractal dimensions 
(with replicas, supersymmetry and Liouville field theory) 
are reliable \cite{CMW} in the regime $|e|<e_c$.
Second, the entropy density for any GREM is known to be self-averaging
in the thermodynamic limit 
\cite{Derrida 1980,Derrida 1985,Derrida 1988,Buffet 1993}
and in view of the close connection between GREM and our problem
we expect this to be also true here.

Thus, although we do not have a rigorous proof, 
we conjecture that $(1/N)\ln\Omega(e)$ is a random variable 
with vanishing variance in the thermodynamic limit $N \to \infty$, 
{i.e.}, self-averaging, and which converges to 
$\lim_{N\rightarrow\infty}(1/N)\ln\langle\Omega(e)\rangle$ . 
This conjecture is supported by the fact that, in addition to our bound, 
it is in this regime that the field theory 
computations of the multifractal dimensions 
(with replicas, supersymmetry and Liouville field theory) are reliable. 
In other words, the entropy should be self-averaging 
in the thermodynamic limit 
and it is given by
\begin{eqnarray}
S(e) & = & \ln \Omega(e)= \langle \ln \Omega(e) \rangle =
\ln \langle \Omega(e)\rangle
\label{eq:S_self_av} \\
& = & N\left(1-\frac{e^2}{e^2_c}\right)\ 
\end{eqnarray}
for any realization of disorder.  Hence, for this
range of energies $|e| < e_{c}$ we may commute the order of $\ln$ and
$ \langle \cdots \rangle $, {i.e.}, 
the ``quenched'' and ``annealed'' entropies 
coincide \cite{quenched-annealed}. 

\subsection{Temperature and free energy}
Up to this point we have proven that the thermodynamics of this model 
is identical to that of the random energy model \cite{Derrida 1980}
for $|e|\geq e_c$ and argued that this is also so for $|e|<e_c$.
In order to obtain the free energy as a function of temperature, 
we follow Derrida \cite{Derrida 1980}. In the region that 
contains states ($|e| < e_{c}$), we obtain the temperature from
\begin{equation}
T = \left( \frac{d s}{d e} \right)^{-1} = -\frac{e^2_c}{2 e}.
\end{equation}
In this region the temperature will be in the range $|T| > {e_c}/{2}$,
and the free energy, computed from the relation ${d f_{0}}/{d T} = -s$,
is given by
\begin{equation}
f_{0}(T) = - T - \frac{e^2_c}{4 T}.
\end{equation}
In the limit when $T \rightarrow T_{c} \equiv {e_{c}}/{2}$, we obtain 
$e \rightarrow -e_{c}$, $s \rightarrow 0$ and $f_{0} \rightarrow -e_{c}$.
Below this temperature, the system cannot lower its energy because there
are no states for $e < -e_{c}$. The system remains frozen
at $e = -e_{c}$, with entropy $s = 0$ and free energy  
$f_{0}(T) = -e_{c}$. In other words, for $|T|$ lower than the freezing 
temperature $T_{c}$, the energy,
the entropy and the free energy remain at their accumulation points. 
In summary,
\begin{equation}
\beta f_{0}(\beta) = 
\cases{
-\left( 1 + \frac{\beta^{2}}{q_{c}^{2}} \right), 
                          & \mbox{if $|\beta| \leq q_{c}$},\cr
- 2 \frac{|\beta|}{q_{c}},      
                          & \mbox{if $|\beta| > q_{c}$},\cr
}
\end{equation}
where $q_{c} \equiv \sqrt{{2 \pi}/{g}}$.

The behavior in the low temperature regime becomes clear if we consider
the $|\beta| \rightarrow \infty$ limit. For any configuration and for a 
given value of $L/a$, there is an absolute minimum $E_{\rm min}$ and an 
absolute maximum  $E_{\rm max}$  for the values of the $V({\bf x})$. For 
large enough $|\beta|$ these extreme values will dominate the partition
function, and one obtains, in the case of positive sign for $\beta$:
\begin{equation}
Z(\beta) \sim e^{- \beta E_{\rm min}}.
\end{equation} 
This means that 
\begin{equation}
\lim_{N \rightarrow \infty}{f_{0}(\beta)} 
= \lim_{N \rightarrow \infty}{\frac{E_{\rm min}}{N}} = -e_{c}, 
\end{equation}
because we know that for a large enough system ${E_{\rm min}}/{N} = -e_{c}$
with probability one. This implies that $f_{0}(\beta)$ is 
self-averaging and equal to $-e_{c} = -2/q_{c}$, as shown above.  

Notice, however, that in the low temperature regime quenched and annealed
values for the free energy are different: 
$ \langle \ln{Z(\beta)} \rangle  \neq \ln{ \langle Z(\beta) \rangle }$. 
Indeed, combining 
Eqs. (\ref{eq:can_micro}) and (\ref{eq:micro_E}) one obtains
for all $\beta$ the so-called lognormal spectrum
\begin{equation}
\ln{ \langle Z(\beta) \rangle } = N \left(1 + 
\frac{\beta^{2}}{q_{c}^{2}}\right), 
\label{eq:F_beta} 
\end{equation}
which shows no obvious sign of the phase transition.  

On the other hand, in the high-temperature regime, the quenched and annealed 
free energies $ \langle \ln{Z(\beta)} \rangle $
and $\ln{ \langle Z(\beta) \rangle }$ do coincide. This can be understood by 
computing
$\ln{Z(\beta)}$ for one particular realization of the disorder
[see Eq.~(\ref{eq:can_micro})]:
\begin{equation}
Z(\beta) = N \int_{\frac{E_{\rm min}}{N}}^{\frac{E_{\rm max}}{N}} de \;\;
e^{ - N\left[ \beta e - s(e)\right]}.
\label{eq:Z(beta)high T}
\end{equation}
If one accepts our conjecture that the entropy density is self-averaging
in the thermodynamic limit, then the saddle point approximation
on the integrand in Eq.~(\ref{eq:Z(beta)high T})
probes the parabolic part of the entropy, and one
obtains for $\ln{Z(\beta)}$ the result of Eq.~(\ref{eq:F_beta}). 
In other words, not only is
$\ln{Z(\beta)}$ self-averaging but in this case $\ln$ and $ \langle \cdots 
\rangle $
commute. In the language of multifractality, this is the region in
which the parabolic approximation is exact.
Besides, since at $|\beta| = q_{c}$ the minimum of $\beta e -s(e)$ falls
at the edge of the populated region, there is a singularity in the 
derivative of $\ln{Z(\beta)}$, and one recovers the phase 
transition point.  

\subsection{Equivalence with random energy models}
\label{sec:rem}

We have calculated the thermodynamic functions of our model 
and they are identical to those of a special
generalized  random energy model
\cite{CMW}.

It is interesting to see how this comes about. All the
results we obtained really follow from our calculation of 
$ \langle \Omega(E)\rangle $
in Eq.~(\ref{eq:micro_sum}). The analogous calculation
for the generalized random energy model of Derrida and Spohn 
\cite{Derrida 1985,Derrida 1988}
gives the same result,
except that positions ${\bf x}$ in the lattice are replaced by
directed paths ${\cal P}$ in a Cayley tree, and the value 
$G({\bf x},{\bf x}) = ({g}/{2 \pi}) \ln(\frac{L}{a})$ is replaced by 
$G({\cal P},{\cal P}) = ({g_{t}}/{\ln{K}}) \ln{d({\cal P},{\cal P})}$.
Then, the two average partition functions must correspond if the parameters
are chosen properly.  
Note that the requirement that $G({\cal P},{\cal P})$ depends logarithmically
on the ultrametric distance $d({\cal P},{\cal P})$ on the tree uniquely defines
the GREM in the thermodynamic limit.

One can take this analysis further by studying moments of the 
microcanonical partition function. Although at first sight the expressions
for the two models appear different, if one changes variables into 
logarithms of the distances, one can actually see that the expressions
for the GREM actually give the Riemann sums that correspond to the 
integrals in the case of our model.

\section{Consequences for the multifractal spectrum}
\label{sec:spectrum}

Now that we have the values of the thermodynamic functions
$s(e)$ and $f_{0}(\beta)$, we can translate them into the 
language of multifractality, using Eqs.~(\ref{eq:tau_can}) and 
(\ref{eq:f_s}) as our dictionary. 

The results are as follows: starting by the 
spectral weight function $f(\alpha)$, we find that it
is defined only in the 
interval $d_{-} \le \alpha \le d_{+}$ (corresponding to the 
entropy being defined only in the interval 
$-e_{c} \le e \le e_{c}$ ), and has the value
\begin{equation}
f(\alpha) = 8 \frac{(d_{+} - \alpha) (\alpha - d_{-})}
{(d_{+} - d_{-})^{2}} .
\label{eq:f_alpha_d}
\end{equation}
However, the values of $d_{-}$ and $d_{+}$ will change with 
the strength of the disorder.

Depending on the strength of the disorder, there are two regimes:
in the weak disorder regime, which corresponds to $g < 2 \pi$,
the quenched and annealed averages for the logarithm of the 
wave function normalization factor $Z(1)$ are coincident,
while in the strong disorder regime, which corresponds to $g > 2 \pi$, 
they are not equal anymore.

In the weak disorder regime,  
the extremal dimensions $d_{-}$ and $d_{+}$ are both positive:
\begin{equation}
d_{\pm} = 2\left(1 \pm \sqrt{\frac{g}{2 \pi}}\right)^{2};
\end{equation}
and $\tau(q)$ has the form 
\begin{equation}
\tau(q)= 
\cases{
2(q-1) \left(1-\frac{q}{{q_{c}}^{2}}\right)
                          & \mbox{if $|q| \leq q_{c}$},\cr
2q \left(1 - \frac{{\rm sgn}(q)}{q_{c}} \right)^{2}     
                          & \mbox{if $|q| > q_{c}$}.\cr
}
\label{eq:tau_weak}
\end{equation}

On the other hand, in the strong disorder regime, 
the lower extremal dimension is zero:
\label{eq:d_weak}
\begin{eqnarray}
d_{-} & = & 0 \nonumber \\
d_{+} & = & 8 \sqrt{\frac{g}{2 \pi}},
\label{eq:d_strong}
\end{eqnarray}
and $\tau(q)$ has the form
\begin{equation}
\tau(q)= 
\cases{
-2q \left(1 - \frac{q}{q_{c}} \right)^{2}     
                          & \mbox{if $|q| \leq q_{c}$},\cr
\frac{4}{q_{c}}(q-|q|)
                          & \mbox{if $|q| > q_{c}$}.\cr
}
\label{eq:tau_strong}
\end{equation}
Notice that in this regime we find $\tau(q) = 0$ for $q > q_{c}$,
meaning that for all integer moments the inverse participation ratio
does not scale with system size. This is usually interpreted as
characteristic of a localized wave function.

\section{Conclusions}
\label{sec:conclusions}

We have calculated the multifractal scaling exponents of the 
critical wave function for two dimensional Dirac fermions in the 
presence of a random magnetic field. 
There is a transition in the multifractal spectrum, 
which is interpreted as a freezing
transition common to glassy systems. 
This freezing transition is
a rigorous result of the present work and had been previously conjectured
on the basis of a comparison with GREM \cite{CMW}.

We have proven that a previously proposed mapping \cite{CMW} between these 
multifractal properties and the thermodynamics of a generalized random
energy model describing directed polymers on a Cayley tree 
\cite{Derrida 1985,Derrida 1988}
is indeed exact in the glassy regime. 
Our proof generalizes entropy considerations on
GREM (Ref.~\cite{Derrida 1980}) to a two dimensional Gaussian field theory.

Derrida has also shown that a direct computation of the quenched free energy
was possible on GREM. This suggests that the same could be done on
the field theory. In fact, it can be shown that the generalization of
Derrida's calculation for GREM naturally leads to estimating
the partition function in Liouville field theory. We thus believe
that there exists a counterpart to the freezing transition of GREM
in Liouville field theory. 
It is an interesting question to probe this issue further.

Another open issue is the fate of replica symmetry if the replica approach
is used to calculate the multifractal scaling exponents. Indeed, 
it is known that the freezing transition in GREM is associated to 
replica symmetry breaking \cite{Derrida 1980,Gross 1984}.
It would be interesting to see how this replica symmetry breaking
manifests itself in a replicated version of our Gaussian field theory.

\section{Acknowledgments}
\label{sec:ack}

We are indebted to J.~Chalker, L.~Balents, A.~W.~W.~Ludwig, M.~P.~A.~Fisher,
and X.-G. Wen 
for many useful comments and penetrating questions. 
This work was completed while three of us (CCC, CM and EF) 
were participants at the Program on 
{Quantum Field Theory in Low Dimensions} 
at the Institute for Theoretical Physics of the 
University of California at Santa Barbara. 
We are very grateful to the organizers of the program and particularly 
to Professor~Jim Hartle, Director of ITP, for his warm hospitality.
This work was supported in part by the NSF through the Grants Nos.
NSF DMR94-24511 and NSF DMR-89-20538
at the University of Illinois at Urbana-Champaign (H.C., C.C.C., E.F. and P.G.),
NSF DMR94-11574 
at Massachusetts Institute of Technology (C.M.),  
and by NSF PHY94-07194 
at the Institute for Theoretical Physics of the University of California at 
Santa Barbara. 

\appendix

\section{Symmetry and Green functions}
\label{sec:GF}

The probability distribution for the disorder,
Eq.~(\ref{eq:distrib}), allows for an exact symmetry 
under a constant real shift of the field configuration 
$\phi({\bf x})$:
\begin{eqnarray}
\phi({\bf x}) & \rightarrow & \phi({\bf x}) + \zeta,
\label{eq:symdef} \\
\psi({\bf x}) & \rightarrow & e^{-\zeta} \psi({\bf x}),
\end{eqnarray}
which leaves both $ P[\phi({\bf x})] $ and $ \Psi({\bf x}) $ unchanged.
However, neither $Z(q)$ nor $\Omega(E)$ are invariant:
\begin{eqnarray}
\Omega(E) & \rightarrow & \Omega(E-2\zeta),  \\
Z(q) & \rightarrow & e^{-2 q \zeta} Z(q), \\
{ \langle }\ln{Z(q)}{ \rangle } & 
\rightarrow & -2 q \zeta +  \langle \ln{Z(q)} 
\rangle . 
\end{eqnarray}
This means that, although $ \langle \ln{Z(q)} \rangle  - q  \langle \ln{Z(1)} 
\rangle $ is well
defined and, in principle, one can compute $\tau(q)$, in a naive 
calculation $ \langle \ln{Z(q)} \rangle $ would be ill defined.

To perform an actual calculation it is convenient to break this 
symmetry in a controlled way.
The simplest approach is to add a mass term $m = 1/L$ to the action. This
penalizes configurations for which $\phi \neq 0$. Another possibility
is to impose Dirichlet boundary conditions on $\phi({\bf x})$. Although
Green functions for $\phi({\bf x})$ in each of these cases will be
different, the only value we need for our purposes is their
short-distance limit, which is the same in all cases, namely,
\begin{equation}
G({\bf x},{\bf y}) \approx -{g\over4\pi}
\ln\left({|{\bf x}-{\bf y}|^2 + a^2 \over L^2}\right).  
\label{eq:green_short}
\end{equation}

\section{Moments of $\Omega(E)$ and $Z(\beta)$}
\label{sec:some moments}

In this appendix, we estimate moments of the density of states $\Omega(E)$
and of the partition function $Z(\beta)$ which, we recall, are related by
Eq.~(\ref{eq:can_micro}). Such moments are needed to decide if
quenched and annealed averages are equal. We begin with $\Omega(E)$.
Let $n$ be an integer larger than one.
By definition,
\begin{eqnarray}
&&
\langle\Omega^n(E) \rangle\equiv
\label{eq:mean n power Omega}
\\
&&
\lim_{W\rightarrow0}
\int D[\phi({\bf x})] P[\phi({\bf x})]
\sum_{{\bf x}_1}\cdots\sum_{{\bf x}_n}
\int{d\lambda_1\over2\pi}\cdots\int{d\lambda_n\over2\pi}
\nonumber\\
&&\times
\exp\left[
{-{W^2\over2}\sum_{k=1}^n\lambda^2_k+iE\sum_{k=1}^n\lambda_k-
2i\sum_{k=1}^n\lambda_k\phi({\bf x}_k)}
\right].
\nonumber
\end{eqnarray}
We assume that all sums and integrals can be freely interchanged.
Averaging over disorder is a Gaussian integral yielding
\begin{eqnarray}
\langle\Omega^n(E)\rangle&&=
\lim_{W\rightarrow0}
\sum_{{\bf x}_1}\cdots\sum_{{\bf x}_n}
\int{d\lambda_1\over2\pi}\cdots\int{d\lambda_n\over2\pi}
\label{eq:n power Omega: integration over disorder}
\\
&&\times\!
\exp\!\left[
{-{1\over2}\!\sum_{k,l=1}^n\!
\lambda_k(W^2\delta_{kl}+4 G_{kl})\lambda_l
+iE\sum_{k=1}^n\lambda_k}
\right]\!.
\nonumber
\end{eqnarray}
Here, $G_{kl}$ is a shorthand notation for the Green function
in Eq.~(\ref{eq:green_short}) with arguments ${\bf x}_k$ and ${\bf x}_l$.
We notice that the integrand on the right-hand side of 
Eq.~(\ref{eq:n power Omega: integration over disorder})
does not depend on ${\bf x}_1,\ldots,{\bf x}_n$ 
for $n=1$ but does for $n>1$. For higher moments than $n=1$,
the statistical correlations encoded by $G_{kl}\neq 0$ for $k\neq l$,
imply that, in a finite system, 
$\langle \Omega^n(E) \rangle\neq\langle \Omega(E) \rangle^n$.
In the thermodynamic limit, the difference between 
$\langle \Omega^n(E) \rangle$ and $\langle \Omega(E) \rangle^n$
disappears if the statistical correlations are short range,
as is the case in the original random energy model (REM) of Derrida,
where all energies are assumed to be identically and independently distributed 
random variables \cite{Derrida 1980}. This is not so, however, if $G_{kl}$
encodes long range statistical correlations as is the case here.
For any moment $n\geq1$, there will be a critical energy density
$e_c(n)\leq e_c(n-1)$ such that the ratio 
$\langle \Omega^n(E) \rangle/\langle \Omega(E) \rangle^n$
diverges in the thermodynamic limit
for $|e|>e_c(n) $. In other words, $\Omega(e)$ is broadly distributed
in the thermodynamic limit. Our claim is that it is the limit
$e_c\equiv\lim_{n\rightarrow1} e_c(n)$ that controls the freezing transition,
and not the naive replica limit $\lim_{n\rightarrow0} e_c(n)=\infty$.

Instead of calculating the sequence $e_c(n)$ explicitly, we
calculate the ratio 
\begin{equation}
R_n(\beta)\equiv {\langle Z^n(\beta)\rangle\over\langle Z(\beta)\rangle^n}.
\end{equation}
We find that
\begin{eqnarray}
R_n(\beta)\sim&&
\int {d^2 x_1\over L^2}...{d^2 x_n\over L^2}
\prod_{i<j}\left|{{\bf x}_i-{\bf x}_j\over L}\right|^{-{2g\beta^2\over\pi}}
\nonumber\\
\sim&&
c_n(g\beta^2)+\left(a\over L\right)^{2(n-1)(1-{ng\beta^2\over2\pi})}.
\label{R_n(q)}
\end{eqnarray}
In the thermodynamic limit, the right hand side is a finite number for
$n\leq(2\pi)/(g\beta^2)$ (assuming $n>1$).  In this case $Z^n(\beta)$
fluctuates weakly. But for $n > (2\pi)/(g\beta^2)$, $R_n(\beta)$ diverges, and
thus $Z^n(\beta)$ fluctuates strongly.  There are important consequences
that follow from Eq.~(\ref{R_n(q)}):
\begin{enumerate}
\item 
There exists a {\it sequence of critical} $\beta_n$ given by
\begin{equation}
\beta^2_n={2\pi\over ng}\equiv {\beta^2_c\over n},
\label{q_c}
\end{equation}
below which
$\ln\langle Z^n(\beta)\rangle/N=$ 
$\ln \langle Z(\beta)\rangle^n/N$ in the thermodynamic limit 
$N\rightarrow\infty$.
Remarkably, the {\it same}
sequence of critical moments is shared by the random energy models
studied in Refs.~\cite{Derrida 1980,Derrida 1985,Derrida 1988}. 
\item Caution is needed when using the replica trick
\begin{equation}
\langle\ln Z(\beta)\rangle=
\lim_{n\rightarrow0} {\langle Z^n(\beta)\rangle-1\over n}.
\label{replica trick}
\end{equation}
Indeed, $\langle Z^n(\beta)\rangle$ is not an analytic function of
$n$ due to singularities at $\beta=2\pi/\beta^2g$, and $n=1$, and caution must 
be used when using the replica trick to calculate $\ln Z(\beta)$.
\end{enumerate}
For GREM, $\beta_c$ indicates a phase transition between the regime
$\beta\leq\beta_c$ and the regime $\beta>\beta_c$. For all GREM,
the quenched and annealed free energy are equal if $\beta\leq\beta_c$.
However, quenched and annealed free energy are not equal if $\beta>\beta_c$,
and need not obey the same functional dependency on $\beta$ for different
GREM in this regime of temperatures. 
There are essentially two Gaussian GREM's whose quenched free energy obey
the same functional dependency on $\beta$:
the REM with uncorrelated energies \cite{Derrida 1980} and the GREM with
logarithmic correlated energies\cite{Derrida 1985,Derrida 1988}. 
Both undergo a sharp freezing
transition characterized by a discontinuous
one step specific heat (as opposed to continuous).
Since we have proven that the same freezing transition characterizes
our multifractal scaling exponents, we conclude that they must be
self-averaging and given by an annealed average below the critical moment
$q_c$.

\end{multicols}


\begin{references}
\bibitem{footnote on topo trivial}
In contrast, if there is a non-vanishing total flux of the magnetic field, 
new effects need to be considered. In the presence of such a flux,
the number of zero modes of the Dirac 
operator are determined by the index theorem, which holds regardless of the 
existence of a topologically trivial random component. 
Consequently, the fields 
$\phi$ and $\chi$ acquire topologically nontrivial components associated with
large gauge transformations and the allowed states become chiral eigenstates 
with  chiral eigenvalue determined by the sign of  the magnetic field.
\bibitem{bound}
The argument goes as follows: if $X$ is a random variable that takes 
only nonnegative values, $P\{\cdots\}$ is the 
probability associated 
with it, and $K$ is any positive constant, then one can write:
\begin{eqnarray}
K P\{ X \ge K\} & = & K \int_{K}^{\infty} P\{X = y\} dy \nonumber \\
& \le & \int_{K}^{\infty} y P\{X = y\} dy \nonumber \\
& \le & \int_{0}^{\infty} y P\{X = y\} dy \equiv  \langle  X  \rangle . 
\nonumber
\end{eqnarray} 
The first inequality holds simply using $K \le y$, and the last
equality is just the definition of average value. By taking $K=1$ and
$X = \Omega_{>}(e)$ we obtain Eq.~(\ref{eq:Omega_plus_bound}), and
by taking $K=1$ and $X = \Omega(e)$ we obtain
Eq.~(\ref{eq:P_Omega_bound}).
\bibitem{Heaviside}
To simplify the argument, we are assuming here a Heaviside 
function regularization for $\delta_{W}(E)$.
\bibitem{quenched-annealed}
A quenched average of random variable $X$ consists in taking first the $\ln$
of $X$ and then averaging. An annealed average of random variable
$X$ is to calculate the mean of $X$ and then take the $\ln$.
\end{references}
\end{document}